# Quantum Public-Key Cryptosystem Based on Classical NP-Complete Problem

Li Yang[*]

State Key Laboratory of Information Security (Graduate School of Chinese Academy of Sciences),
Beijing 100039, P. R. China

We present a quantum public-key cryptosystem for quantum message transmission based on a classical NP-complete problem related with finding a code word of a given weight in a linear binary code.

A generator matrix $G$ for a linear code $C$ is a $k$ by $n$ matrix for which the rows are a basis of $C$. The parity check matrix $H$ of $C$ satisfies $GH^T = 0$. For every error $e^{(j)}$, $s^{(j)} = e^{(j)} H^T$ is called the syndrome of $e^{(j)}$. A Goppa code $\Gamma(L, g)$ is a linear error-correcting code that is defined with Goppa polynomial $g(z)$[1]. There is a fast algorithm for decoding Goppa codes, but the general problem of finding a code word of a given weight in a linear binary code is NP-complete. Based on this, similar to that suggested by McEliece in classical cryptology [2] and the method suggested in [3], we present here a quantum public-key cryptosystem with classical computational security based on NP-complete problem.

A quantum message is a sequence of pure states. Without loss of generality, we restrict our attention to the encryption and decryption of a pure state. Suppose $G$ is a generator matrix of a Goppa code, $G' = SGP$, where S is an invertible matrix and P is a permutation matrix. We choose $G'$ as the public key and ($S, G, P$) as the private key. The encoding and decoding process are:

I. Encoding

Alice uses Bob's public key $G'$ to encrypt the state $\sum_m \alpha_m |m\rangle$ as below:

1. Compute:
$$\hat{U}_{G'} \sum_m \alpha_m |m\rangle_k |0\rangle_n = \sum_m \alpha_m |m\rangle_k |mG'\rangle_n,$$

where the symbole $|\ \rangle_i$ represents the state of a register with $i$ qubits, and the

---

[*] E-mail: yangli@gscas.ac.cn

unitary operator $\hat{U}_{G'}$ is defined as

$$\hat{U}_{G'}|m\rangle_k|s\rangle_n = |m\rangle_k|s \oplus mG'\rangle_n ;$$

$$\hat{U}_{G'^-} \sum_m \alpha_m |m\rangle_k |mG'\rangle_n = \sum_m \alpha_m |m \oplus mG'G'^-\rangle_k |mG'\rangle_n = |0\rangle_k \sum_m \alpha_m |mG'\rangle_n ,$$

where the unitary operator $\hat{U}_{G'^-}$ is defined as

$$\hat{U}_{G'^-}|m\rangle_k|s\rangle_n = |m \oplus sG'^-\rangle_k|s\rangle_n ,$$

the matrix $G'^-$ is a generalized inverse matrix of $G'$. Because $G'$ is a full row rank matrix, there exists $G'^-$ that satisfies $G'G'^- = I_k$. This is the condition that one can get $\sum_m \alpha_m |mG'\rangle_n$ from $\sum_m \alpha_m |m\rangle_k$.

2. Compute:
$$\hat{U}_{E^{(i)}} \sum_m \alpha_m |mG'\rangle_n \equiv \sum_m \alpha_m |mG' \oplus e^{(i)}\rangle_n$$
and send it.

II. Decoding

Bob uses his private key $(S, G, P)$ to decrypt the state coming from Alice:

1. Compute:
$$\hat{U}_{P^{-1}} \sum_m \alpha_m |mG' \oplus e^{(i)}\rangle_n \equiv \sum_m \alpha_m |mG'P^{-1} \oplus e^{(i)}P^{-1}\rangle_n = \sum_m \alpha_m |mSG \oplus e^{(j)}\rangle_n .$$

2. Compute:
$$\hat{U}_H \sum_m \alpha_m |mSG \oplus e^{(j)}\rangle_n |0\rangle_{n-k} = \sum_m \alpha_m |mSG \oplus e^{(j)}\rangle_n |s^{(j)}\rangle_{n-k} ,$$

where the unitary operator $\hat{U}_H$ is defined as

$$\hat{U}_H |m\rangle_n |s\rangle_{n-k} = |m\rangle_n |s \oplus mH^T\rangle_{n-k} .$$

3. Measure the second register to get $s^{(j)}$, and find $e^{(j)}$ via the fast decoding algorithm of the Goppa code generated by $G$;

4. Compute:
$$\hat{U}_{e^{(j)}} \sum_m \alpha_m |mSG \oplus e^{(j)}\rangle_n = \sum_m \alpha_m |mSG\rangle_n ,$$

where the unitary operator $\hat{U}_{e^{(j)}}$ is defined as:

$$\hat{U}_{e^{(j)}} |m\rangle_n = |m \oplus e^{(j)}\rangle_n .$$

5. Compute:

$$\hat{U}_{G^-} \sum_m \alpha_m |mSG\rangle_n |0\rangle_k = \sum_m \alpha_m |mSG\rangle_n |mSGG^-\rangle_k$$
$$= \sum_m \alpha_m |mSG\rangle_n |mS\rangle_k,$$

where the unitary operator $\hat{U}_{G^-}$ is defined as:

$$\hat{U}_{G^-} |m\rangle_n |s\rangle_k = |m\rangle_n |s \oplus mG^-\rangle_k;$$

and compute:

$$\hat{U}_G \sum_m \alpha_m |mSG\rangle_n |mS\rangle_k = \sum_m \alpha_m |mSG \oplus mSG\rangle_n |mS\rangle_k = |0\rangle_n \sum_m \alpha_m |mS\rangle_k,$$

where the unitary operator $\hat{U}_G$ is defined as:

$$\hat{U}_G |m\rangle_n |s\rangle_k = |m \oplus sG\rangle_n |s\rangle_k.$$

6. Compute:

$$\hat{U}_{S^{-1}} \sum_m \alpha_m |mS\rangle_k \equiv \sum_m \alpha_m |mSS^{-1}\rangle_k = \sum_m \alpha_m |m\rangle_k.$$

Then Bob gets the state coming from Alice.

I would like to thank G. L. Long for useful discussions.